\documentclass[prl,reprint]{revtex4-1}
 \usepackage[utf8x]{inputenc}
 \usepackage[T1]{fontenc}
\usepackage{diagbox}
\usepackage{amsfonts}
\usepackage{amsmath}
\usepackage{amsthm}
\usepackage{indentfirst}
\usepackage{microtype}
\usepackage[squaren]{SIunits}
\usepackage{booktabs}
\usepackage{graphicx}
\usepackage{multirow}
\usepackage{wallpaper}

\usepackage{braket}
\usepackage{caption}
\usepackage{subcaption}
 \makeatletter
 \g@addto@macro\normalsize{%
   \setlength\abovedisplayskip{15pt}
   \setlength\belowdisplayskip{15pt}
   \setlength\abovedisplayshortskip{15pt}
   \setlength\belowdisplayshortskip{15pt}
 }

\makeatother

\begin{document}
\title{Optical properties of high pressure liquid hydrogen across molecular dissociation}
\author{Giovanni Rillo$^1$, Miguel A. Morales$^2$,  David M. Ceperley$^3$, Carlo Pierleoni$^{4,5,*}$ }

\affiliation{
$^1$Department of Physics, Sapienza University of Rome, Italy\\
$^2$Physics Division, Lawrence Livermore National Laboratory, California USA\\
$^3$Department of Physics, University of Illinois Urbana-Champaign, USA\\ 
$^4$ Department of Physical and Chemical Sciences, University
of L'Aquila, Via Vetoio 10, I-67010 L'Aquila, Italy\\
$^5$ Maison de la Simulation, CEA, CNRS, Univ. Paris-Sud, UVSQ, Universit\'e Paris-Saclay, 91191 Gif-sur-Yvette, France
}

\date{\today}
\begin{abstract}
Optical properties of compressed fluid hydrogen in the region where dissociation and metallization is observed are computed by ab-initio methods and compared to recent experimental results. We confirm that above 3000 K both processes are continuous while below 1500K the first order phase transition is accompanied by a discontinuity of the DC conductivity and the thermal conductivity, while both the reflectivity and absorption coefficient vary rapidly but continuously. Our results support the recent analysis of NIF experiments (P. Celliers et al, Science 361, 677--682 (2018)) which assigned the inception of metallization to pressures where the reflectivity is about 0.3. Our results also support the conclusion that the temperature plateau seen in laser-heated DAC experiments at temperatures higher than 1500 K corresponds to the onset of of optical absorption, not to the phase transition.
\end{abstract}

\maketitle
Metallization of hydrogen is a fundamental yet elusive process both in the crystalline and liquid phases. 
Metallic hydrogen has been unequivocally detected by Weir et al. in the liquid phase at pressures of $\sim 140$GPa and temperatures of 2500-3000K using dynamical compression with shock waves\cite{Weir1996,Nellis1999}. 
The emergence of the metallic state was detected by a direct measure of sample resistivity, but they were unable to make a clear characterization of the insulator-metal (IM) transition \cite{Nellis1998}.
\begin{figure*}[ht]
\includegraphics[width=0.7\textwidth]{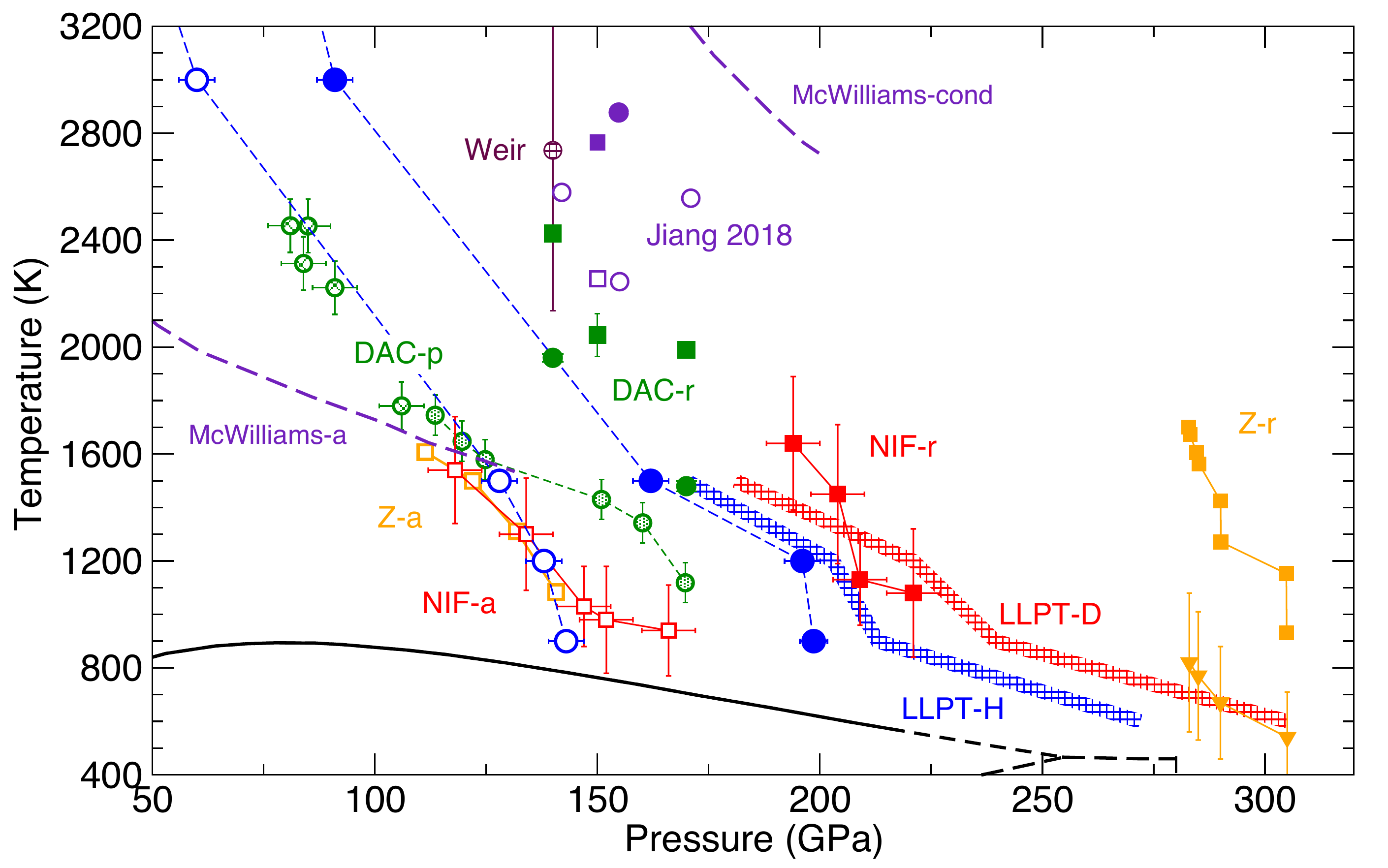}
\caption{\small{Phase diagrams of hydrogen and deuterium around the LLPT line. Shaded lines (blue for hydrogen and red for deuterium)  are the LLPT predicted by CEIMC \cite{Pierleoni2016,Pierleoni2018a}. Closed symbols are estimates of the LLPT from the reflectivity coefficient; open symbols indicate the inception of absorption. Squares correspond to deuterium, circles to hydrogen. Shown are data from sp-DAC (green), from Z-machine (orange), from NIF (red) and from lp-DAC methods (purple).  
  DAC-r: data from sp-DAC at R=0.3; Z-r: data from Z-machine at the observed discontinuity in reflectivity (squares original data points, downward triangles as reanalyzed in ref \cite{Celliers2018}; NIF-r: data from NIF at R=0.3. DAC-p: data from sp-DAC corresponding to the temperature plateau ref. \cite{Dzyabura2013,Zaghoo2016} ($T\leq 1700$) and ref. \cite{Ohta2015} ($T\geq 1700$); Z-a: data from Z-machine when the sample becomes dark; NIF-a: data from NIF when the absorption coefficient exceeds 1$\mu m^{-1}$; lp-DAC \cite{Jiang2018}:  closed purple points are conducting conditions and open purple points are non-conducting conditions (for both hydrogen and deuterium). Two dashed purple lines indicate the inception of absorption (McWilliams-a) and the metallic boundary (McWilliams-cond) \cite{McWilliams2016}. Brown shaded circles (Weir) show the inception of metallicity from gas gun experiments \cite{Weir1996}. Blue points are theoretical estimates from this study: closed circles show when $R=0.3$ for H/vacuum interface; open circles when the absorption coefficient equals $1\mu m^{-1}$. }}
 \label{fig:phase_diagram}
\end{figure*}
Recent static compression experiments, using a Diamond Anvil Cell (DAC) with controlled laser heating (short-pulses DAC, sp-DAC)\cite{Silvera2009,Dzyabura2013,Zaghoo2016,Zaghoo2017a,Zaghoo2018a}, studied liquid hydrogen and deuterium measuring both temperature and pressure. An absorber was heated with short laser pulses; the heat was transferred to the sample by thermal conduction. %, both being pressurized inside a DAC. 
The sample temperature was observed to grow linearly with the power of the laser impulses until a plateau in the temperature was observed. The onset of the plateau was interpreted as the occurrence of a first-order phase transition \cite{Dzyabura2013}. 
Similar experiments at higher temperatures (T>2000 K) confirmed the observation \cite{Ohta2015}. %Complementing the experiments with optical measurements is crucial to investigate the occurrence of the metallic state. 
Optical measurements during laser heating observed an increase in reflectivity  until saturation when $R\simeq 0.5$, at a temperature higher than the plateau\cite{Zaghoo2016,Zaghoo2017a,Zaghoo2018a}. See fig. (1).

During shock (ramp) compression experiments on deuterium using the Z-machine \cite{Knudson2015},  the sample became first opaque and later, at higher pressure, have an abrupt change in reflectivity, suggesting a discontinuous IM transition. However, this transition occurred at a much higher pressure ($\sim$150GPa) than in the DAC experiments. Temperature was not directly measured; it was inferred using a model Equation of State (EOS). They found a temperature independent transition line in contrast to the observation for hydrogen in the DAC experiments. Such a large difference cannot be ascribed to the isotopic effect.
Previous shock compression experiments on deuterium by Fortov et al.\cite{Fortov2007} found indirect evidence of a discontinuous transition much closer to those of Zaghoo et al. \cite{Zaghoo2016}.
Very recent experiments with shock compression of deuterium at NIF confirmed the Z-machine observation of a first regime of absorbtion followed by a rapid rise of reflectivity, again up to a plateau of $\sim 0.5$. However, this rise was observed at lower pressures, roughly 100GPa lower than at the Z-machine, closer to the results of static compression experiments and in striking agreement with theoretical predictions from Quantum Monte Carlo methods \cite{Morales2010,Pierleoni2016,Mazzola2018}. The metallization transition was assumed to occur when reflectivity at the D-LiF interface equaled $0.3$. This corresponds to the minimum metallic conductivity of $\sim 2000 S/cm$ \cite{Weir1996,Nellis1999}. Contrary to the Z-machine experiments they did not observe hysteresis in the reflectivity during compression and decompression of the sample. 

In a third experimental method, a $\mu s$ long laser impulse heats a DAC  sample (lp-DAC) \cite{McWilliams2016,Jiang2018}. Using ultrafast spectroscopy, both the signature of high temperature metallization and the emergence of the absorbing state during the cooling process was detected. % before recovering the transparent character of the sample at even lower temperature. 

In fig. \ref{fig:phase_diagram} we show the emerging phase diagrams of hydrogen and deuterium from the various experiments. Also shown are the liquid-liquid phase transitions  (LLPT)  for the two isotopes from Coupled Electron-Ion Monte Carlo calculations\cite{Pierleoni2016} as well as new results obtained in this work from optical properties. 

Note that shock compression experiments do not provide direct information about the molecular character of the sample and it is very difficult to observe the weakly first-order character of the liquid-liquid transition \cite{Pierleoni2016}. % during the pressure history of the experiments. 
The signal of the transition is obtained from optical measurements, mainly the reflectivity and absorption coefficients. Moreover, in shock experiments, the  temperature is inferred from theoretical models which can lead to large uncertainties. In static compression experiments information about the molecular character can be inferred by vibrational spectroscopy. This, however, is difficult near the transition once the reflectivity increases. 

\begin{figure*}
  \includegraphics[width=\textwidth]{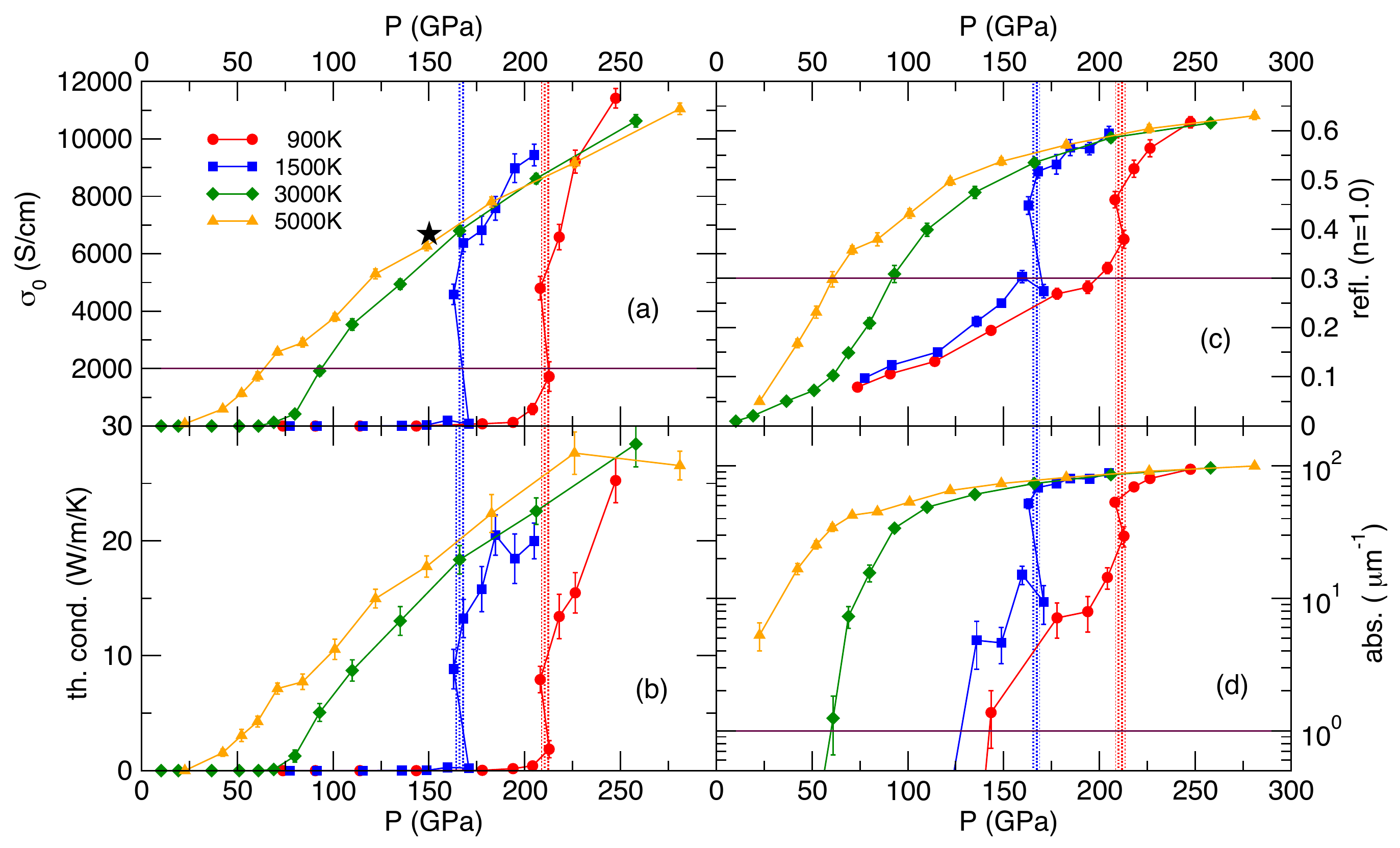}
  \caption{\small{Linear response properties along the isotherms: T = 900K, 1500K, 3000K and 5000K. Configurations along the two lower isotherms are obtained by CEIMC, while configurations at 3000K and 5000K are obtained by PIMD. Panel (a): Static electrical conductivity $\sigma_0$ . The black star is a data point from ref \cite{Jiang2018} for deuterium at 4400K. Panel (b): thermal conductivity.
Panel (c): reflectivity  at $\omega=2.3$ eV corresponding to an optical wavelength $\lambda=539$ nm, at a vacuum  interface.
Panel (d): absorption coefficient at $\omega=2.3$ eV. The vertical dashed lines indicate the pressure of the LLPT as observed in the EOS\cite{Pierleoni2016} for T=900K (red) and  T=1500K (bue). The horizontal line in panel (a) represents the minimum metallic conductivity of $2000 S/cm$, while in panels (c) and (d) corresponds to the threshold values used in interpreting the NIF experiments \cite{Celliers2018}.
  }}
  \label{fig:sigma0}
\end{figure*}    

Theoretical predictions of hydrogen metallization and molecular dissociation have been provided by chemical models\cite{Ebeling1985,Saumon1989}, which suggested the existence of a first-order transition line, and by first-principle simulation methods\cite{McMahon2012a}. Results of early calculations \cite{Magro1996,Scandolo2003,Bonev2004,Delaney2006a,Vorberger2007a,Vorberger2007b,Holst2008} gave differing predictions. %whether molecular dissociation would occur through a cross-over or through a phase transition. 
More recent and accurate investigations \cite{Morales2010,Liberatore2011,Lorenzen2010,Holst2011,Morales2013liquid,Knudson2015,Pierleoni2016,Norman2017a,Mazzola2018} indicate the presence, below a critical temperature, of a first-order transition between an insulating molecular fluid and a conducting monoatomic fluid. This picture emerges both from Born-Oppenheimer Molecular Dynamics (BOMD) with several exchange-correlation approximations and from Coupled Electron-Ion Monte Carlo (CEIMC) \cite{Pierleoni2006}. Hence, the existence of a first-order LLPT with a negative P-T slope is a robust prediction of theory. Nuclear quantum effects change the location of the phase transition by tens of GPa \cite{Morales2013liquid,Knudson2015,Pierleoni2016} but preserve the first-order character. The location of the transition line, on the other hand, depends significantly on the details of the simulation. The results from Quantum Monte Carlo based methods (CEIMC \cite{Pierleoni2016} and Quantum Monte Carlo Molecular Dynamics (QMC-MD)\cite{Mazzola2018}), more reliable than BOMD, lie between the static compression experiment predictions \cite{Ohta2015,Zaghoo2016} and the dynamic compression from Z-Machine \cite{Knudson2015} and are in excellent agreement with the recent NIF data \cite{Celliers2018}. %The CEIMC line also lies in between the BOMD lines from various approximations, closer to the line from the vdW-DF approximation, hence confirming the superiority of this functional to study hydrogen\cite{Clay2014}.

Electronic properties across the transition region, such as optical conductivity and reflectivity, can be computed by the Kubo-Greenwood formula \cite{Kubo1957a,Greenwood1958} within DFT \cite{Morales2010,Lorenzen2010,Holst2011,Morales2013liquid,Pierleoni2016}. \footnote{A correlated-electron method (Correlation Function QMC method\cite{Lin2009}) to compute electrical conductivity beyond the single electron picture of Kubo-Greenwood has only been applied at temperatures much higher than the critical point.}
Using nuclear configurations sampled by both BOMD and CEIMC, the static conductivity is found to be discontinuous at the transition. The molecular character of the fluid suddenly disappears within CEIMC \cite{Pierleoni2016,Pierleoni2018a}, hence the dissociation transition coincides with the insulator-metal transition. At the same time, a discontinuous behavior of the electronic momentum distribution is observed at the transition and is associated with a change of electronic localization from non-Fermi to Fermi liquid character\cite{Pierleoni2018b}. Within BOMD, the molecular character disappears more slowly above the transition, but details depend on the specific functional employed \cite{Morales2010,Knudson2015}.

In this work we use CEIMC and Path Integral Born-Oppenheimer MD (PIMD) to perform simulations of high-pressure liquid hydrogen in the region of the molecular dissociation and metallization in a temperature range including both the first-order transition at low temperature and the cross-over at higher temperature. We then compute the optical properties with DFT, in particular the reflectivity and the absorption coefficient which are the key measured properties \footnote{The experimental determination of conductivity has been achieved only in early gas gun experiments\cite{Weir1996,Nellis1999}. }.
We find that even at temperatures below the critical point, where the DC conductivity has a discontinuity of $\sim 4$ orders of magnitude, the reflectivity at the experimental frequency shows only a rapid increase, in agreement with the NIF observations and at variance with the Z-machine data. 
In agreement with the experimental picture, we find a lower pressure for the sample to become absorbing (assigned to the pressure at which absorption equals $1\mu m^{-1}$) and a higher pressure for the reflectivity to exceed $R\simeq 0.3$. Moreover, we show that below the critical point, where the variation in reflectivity across the LLPT is more rapid, the value $R\simeq 0.3$ corresponds to a pressure value very close to the observed LLPT, and hence to the transition to the metallic state. Above the critical point, this criterium ($R\simeq 0.3$) is more qualitative but still rather a good indication of the crossover. 
We observe a substantial reflectivity even for ``insulating'' molecular hydrogen at conditions close to the transition line.  Our calculations show that the rapid but continuous change of reflectivity observed in dynamic experiments \cite{Celliers2018} is consistent with a first-order liquid-liquid transition. 
Moreover, our calculations suggest that, for T$\geq$1500 K, the temperature plateaux observed in sp-DAC experiments  \cite{Dzyabura2013,Ohta2015,Zaghoo2016} corresponds to the inception of optical absorption, not to the phase transition.

CEIMC simulations of 54  and 128 hydrogen atoms were performed at densities corresponding to the range of pressure between 50GPa and 250GPa along three isotherms: T=900, 1200, 1500 K.  Details of the calculations and a careful assessment of their accuracy are reported in the Supporting Information of ref. \cite{Pierleoni2016}. 
PIMD simulations were performed  along the isotherms T=3000, 5000, 6000, 8000 K. We used a customized version of VASP employing the vdW-DF1 exchange correlation functional, which has proven to have the best energetics among different computationally affordable functionals when benchmarked by Quantum Monte Carlo\cite{Clay2014}. It also provides reasonably accurate results in low temperature molecular H$_2$ crystals \cite{Rillo2017a}. Both cells with 54 and 128 atoms were used with a 3x3x3 k-grid and $\Gamma$ point sampling, respectively. 
%The number of proton time %at T=1500 and 2400 K at $\sim$141 GPa to compare with existing experiments and additional simulations
%slices in the Path Integral set to be $N_p \ge 9000K/$T.
%the number of proton slices was set to $N_P=4$ for T=3000 and 5000 K and $N_P=2$ for T=6000 and 8000 K.
Linear response theory (Green-Kubo) using the HSE functional with 25\% of exact-exchange\cite{Heyd2005} was used to calculate the dielectric response.  
At each density and temperature, 16 snapshots of the nuclear positions were extracted from the equilibrium trajectory. For each snapshot we computed optical conductivity, reflectivity, absorption coefficient and electronic thermal conductivity, related to the power absorbed by the system and to its heat conduction. The properties were then averaged over the snapshots to obtain the statistical average.
Static ($\omega=0$) values of the conductivities are obtained by extrapolating finite energies ($\omega>0$) data using a standard procedure \cite{Morales2010,Holst2011}. 

%;$N_P=6$ for T=1500 K and $N_P=4$ for T=2400 K.
%  One legitimate question is how finite size effects and vdW-DFT influence our results. Without claiming to perform an exhaustive study, fig. 
%    \ref{fig:stewart} shows that:
%    \begin{itemize}
%      \item the agreement between conductivities when different cell sizes is good at high energies, and gets worse  around $\omega=0$; however, our static values
%        are based on extrapolation from higher energies, so that turns out to be irrelevant. 
%      \item the agreement between CEIMC and vdW-DFT is overall quite good: moreover, the CEIMC N=54 conductivity is very close to the vdW-DFT N=128 one, suggesting
%        that finite size effects have been effectively taken care of.
%      \item There is  a discrepancy  in pressure of about 15 GPa between CEIMC and vdW-DFT. It is difficult to say if this  can be assumed to be a 
%        systematic shift
%    \end{itemize}

%\section{Results and discussion} \label{sec:results}
Our main results are reported in fig. \ref{fig:sigma0}.
The electrical and thermal conductivities are static values ($\omega=0$); the reflectivity and the absorption coefficient are computed at $\omega=2.3$ eV, corresponding to $\lambda=539$nm, a typical value used in experiments \cite{McWilliams2016,Zaghoo2018a,Knudson2015,Celliers2018}. 
The left panels clearly show a discontinuity in the electrical and thermal conductivities at lower temperatures, indicating a first-order IM transition.
The curves are smooth at higher temperatures ($T\geq 3000K$), consistent with the termination of the first-order transition line at the estimated critical temperature (between 1500K and 3000K)  \cite{Morales2010,Pierleoni2016,Pierleoni2018a}. Above the critical temperature, molecular dissociation and metallization become a continuous ``cross-over''.
%Incidentally, the highest computed conductivity ($\simeq 12000 S/cm$) would put liquid hydrogen on par with steel, hence to a poor metal.

In contrast, the reflectivity and absorption coefficients, shown in the right panels of fig. \ref{fig:sigma0}, are not discontinuous, even below the critical point. Absorption coefficients larger than 1$\mu m^{-1}$, the threshold value employed in ref. \cite{Celliers2018} between transparent and opaque regime, happens between 50 and 150 GPa depending on temperature. Those pressure values are shown in fig. \ref{fig:phase_diagram} as open blue circles. The pressure difference between this value and the critical pressure is about 50GPa. The absorption saturates at $\simeq 100\mu m^{-1}$ at higher pressures. The darkening of the sample with pressure is associated with the reduction of the energy gap to the energy of the laser's photons. 
Reflectivity also increases smoothly with pressure from zero up to a saturation value of $\sim 0.6$. For temperatures below the critical point we observe an abrupt jump of $\sim 0.1-0.2$ in the reflectivity in close correspondence to the LLPT transition (more visible at T=1500K). However, the reflectivity at pressures just below the transition is already quite substantial (R$\sim 0.3$). The horizontal line in panel (c) represents the threshold value $R=0.3$ employed in ref. \cite{Celliers2018} to establish the minimum metallic conductivity of 2000 S/cm, used to estimate the IM transition. The pressures when  $R=0.3$ at various temperatures are shown in fig. \ref{fig:phase_diagram} as closed blue circles. Comparing panels (a) and (c) of figure \ref{fig:sigma0} we see that $R=0.3$ matches well with the pressures where the conductivity reaches 2000 S/cm confirming the experimental analysis \footnote{A similar analysis is reported in the Supplementary Material of ref. \cite{Knudson2015} (see fig. S7).}. Below the critical temperature, the value $R=0.3$ is very close to the observed reflectivity on the low pressure side of the LLPT.  

\begin{figure}[h]
  \includegraphics[width=\columnwidth]{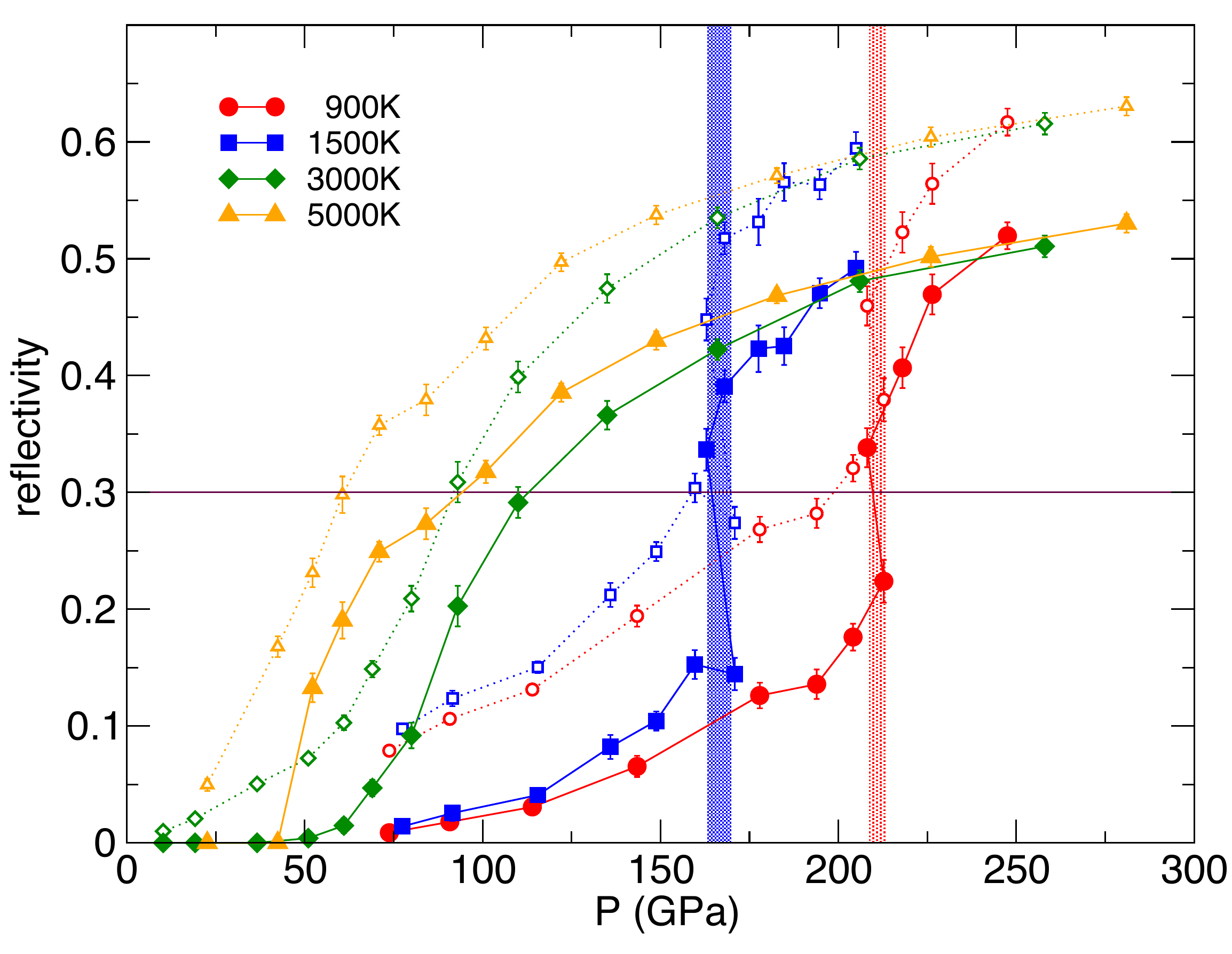}
  \caption{\small{Dependance of the reflectivity along isotherms for differing window materials. Closed symbols correspond to a material with $n=1.49$ (LiF) while open symbols to a material with $n=1.0$ (vacuum). Color and symbols, as well as vertical dashed and horizontal continuous lines, are as in figure \ref{fig:sigma0}.}}
  \label{fig:reflect}
\end{figure}

The reflectivity in Fig. \ref{fig:sigma0}(c) is computed assuming a reflecting interface with air ($n=1.0$). However, in experiments different materials are used. At NIF, a LiF window is employed with a refractive index of $n\sim 1.5$ at the relevant pressures. Figure \ref{fig:reflect} shows the reflectivity obtained with a material of refractive index $n=1.49$ and compares with the data in Fig. \ref{fig:sigma0}(c). We see a rigid downward shift of the reflectivity along all isotherms; the saturation value at high pressures drops from $\sim 0.6$ to $\sim 0.5$. Below the critical point, due to the vertical jump of reflectivity at the transition, the pressure corresponding to R=0.3 is slightly increased improving the agreement with the LLPT pressure, which confirms the validity of the criterium used to interpret the NIF experiments. However, above the critical temperature changing the refractive index of the contrasting medium substantially changes the pressure corresponding to $R=0.3$. While for $n=1$,  $R=0.3$ and $\sigma_0=2000 S/cm$ corresponds to the same pressures. With $n=1.49$ the two pressures differ: 20 Gpa at 3000K and 35GPa at 5000K. We note that this criterium is only qualitative. Increasing the refractive index of the window material will require higher pressures to reach R=0.3, corresponding to larger values of the conductivity.

Figure \ref{fig:phase_diagram} shows the estimated hydrogen and deuterium phase diagrams in the region of the liquid-liquid transition both from experiments and theory. There are two main sets of data points aligned along diagonal descending lines. The line at lower pressure is where the absorption coefficient reaches the threshold of $1 \mu m^{-1}$ according to NIF data and to our present results (from figure \ref{fig:sigma0}(d)). Data points from the observed temperature plateaus in sp-DAC experiments follow the same behavior above 1500K. At lower temperatures they move to slightly higher pressure. The second diagonal line is where the reflectivity reaches the value $R=0.3$ in NIF experiments and in our calculations with $n=1.0$. Note that NIF data are for deuterium while our results are for hydrogen, so we do not expect perfect agreement. We also report the LLPT transition line from ref. \cite{Pierleoni2016}. As noted above, the value $R=0.3$ is reached slightly before the transition point. Data at $n=1.49$ would be in better agreement with the transition pressures below the critical temperature and the high temperature point will move toward higher pressure by about 20-30 GPa (and even more for windows with larger refractive index). The reflectivity data from static compression experiments lie along the same line both for hydrogen \cite{Zaghoo2017a} and deuterium \cite{Zaghoo2018a} \footnote{The same criterium of $R=0.3$ has been used to extract these data \cite{Zaghoo_private}.}

We performed simulations of high pressure liquid hydrogen in the region of molecular dissociation and metallization, employing state of the art first-principles methods (CEIMC and vdFW-PIMD). We computed optical properties in the Kubo-Greenwood framework and compare with existing experimental data. Our work confirms the validity of the proposed analysis of NIF data, in particular that a rapid but continuous increase of reflectivity is still compatible with the existence of a weakly first-order LLPT found in first-principles simulations \cite{Pierleoni2016}. Moreover, we confirm that below the critical temperature, the transition pressure can be associated with a reflectivity value of R$\simeq$0.3, which also corresponds to the sudden jump in conductivity of roughly 4 order of magnitudes. Above the critical point, where the dissociation-metallization process is continuous, a reflectivity of R=0.3 corresponds to conductivity values larger than 2000S/cm, the precise value depending on the refractive index of the window material.

 \begin{acknowledgments}
We thank Paul Loubeyre, Isaac Silvera, Mohamed Zaghoo, Ronald Redmer and Peter Celliers for useful discussions. 
M. A. M. acknowledges support from the US DOE, Office of Science, Basic Energy Sciences, Materials Sciences and Engineering Division, as part of the Computational Materials Sciences Program and Center for Predictive Simulation of Functional Materials (CPFSM).
 This work was performed in part under the auspices of the US DOE by Lawrence Livermore National Laboratory under Contract DE-AC52-07NA27344. D.M.C. was supported by DOE Grant NA DE-NA0001789 and by the Fondation NanoSciences (Grenoble). C.P. was supported by the Agence Nationale de la Recherche (ANR) France, under the program ``Accueil de Chercheurs de Haut Niveau 2015'' project: HyLightExtreme. Computer time was provided by PRACE Project 2016143296, by an allocation of the Blue Waters sustained petascale computing project, supported by the National Science Foundation (Award OCI 07- 25070) and the State of Illinois, and by the HPC resources from GENCI-CINES under the allocation 2018-A0030910282.  
\end{acknowledgments}\bibliographystyle{apsrev4-1}

\bibliography{dottorato}

\end{document}